\documentclass[letterpaper,10pt]{article} 

\usepackage{opticameet3} 
\usepackage{fullpage,graphicx,psfrag,url, amsmath,color,cite}

\usepackage{amsmath,amssymb}

\usepackage{readarray}
\usepackage[colorlinks=true,bookmarks=false,citecolor=blue,urlcolor=blue]{hyperref} 
\usepackage{pgfplots} 
\usepackage{pgffor}
\pgfplotsset{compat=1.16}
\usepgfplotslibrary{colorbrewer}
\pgfplotsset{cycle list/Set1-4}

\usepgfplotslibrary{fillbetween}

\tikzset{
    partial ellipse/.style args={#1:#2:#3}{
        insert path={+ (#1:#3) arc (#1:#2:#3)}
    }
}

\begin{document}

\title{Performance limits of spectro-temporal unitary transformations for coherent modulation}


\author{Callum Deakin, Xi Chen}

\address{Nokia Bell Labs, 600 Mountain Ave, Murray Hill, NJ, USA}

\email{callum.deakin@nokia-bell-labs.com} 

\begin{abstract}
We analyse the performance limits of coherent modulation based on lossless unitary transformations, demonstrating that they can achieve high ($>$30~dB) SINAD and outperform conventional IQ modulators at equivalent transmitter laser powers.
\end{abstract}
{\vspace{0.1cm}}
\section{Introduction}

\begin{figure}[b]
\vspace{-0.5cm}
   \centering
    \includegraphics[width=0.8\linewidth]{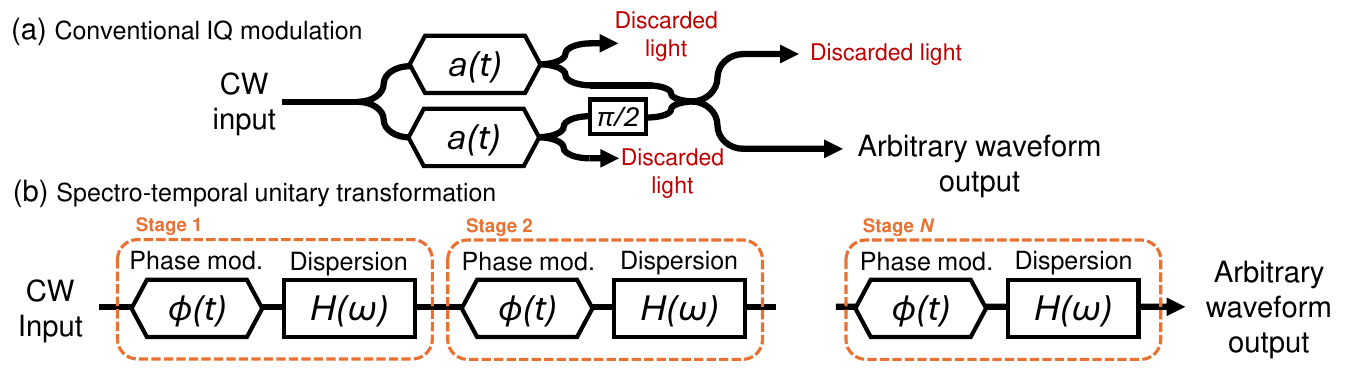}
        \vspace{-0.3cm}
    \caption{(a) Conventional IQ modulation based on amplitude modulation, $a(t)$. (b) Lossless spectro-temporal unitary transform based arbitrary waveform modulation based on cascaded phase modulators $\phi(t)$ and dispersive elements $H(\omega)$.}
    \label{experimenf_setup}
    \vspace{-0.1cm}

\end{figure}

Conventional coherent modulation is based on amplitude modulation via Mach-Zehnder modulators (MZMs) which is inherently lossy since it is based on switching. Excess light is discarded via the unused output port of the optical coupler that combines the two arms of an MZM and the optical coupler that combines the in-phase (I) and quadrature (Q) parts of the optical field, as shown in Fig.~\ref{experimenf_setup}(a). MZM-based coherent transmitters often have modulation losses on the order of 20~dB. Such significant loss is detrimental for power-efficiency and may become prohibitive for future transceivers with THz level bandwidths, such as highly parallel or frequency-comb-based transceivers, where the available power on the photonic integrated circuit is limited. 

An alternative modulation scheme based on multiple stages of alternating phase modulators and dispersive elements (Fig.~\ref{experimenf_setup}(b)) has been proposed and demonstrated~\cite{mazur2019optical,thiel2017programmable}, inspired by its spatial analog: the multi-planar light converter (MPLC)~\cite{morizur2010programmable}. Conceptually, the scheme achieves amplitude modulation by redistributing light temporally rather than switching. This scheme is known as a spectro-temporal unitary transformation or all-pass modulation, as the relation between the output target optical waveform (e.g. a Nyquist-pulse-shaped $N$-QAM waveform) and the input CW light can be viewed as a unitary transformation on the spectro-temporal modes. Since it uses only phase modulators and dispersive elements, it is theoretically lossless. It has been recently shown that this modulation scheme can be operated in real time\cite{saxena2023performance} and can modulate the different wavelengths of a frequency comb independently without any wavelength mux/demux~\cite{mazur2019multi}. Besides classical optical communications, the lossless nature of these transformations also make them highly attractive for quantum information processing~\cite{thiel2017programmable,ashby2020temporal}.

Though experimentally demonstrated, there has been little to no discussion on the performance limits of this modulation scheme. Here, we examine how the key design factors, including number of stages, amount of dispersion, and phase modulator bandwidth impact the accuracy of these spectro-temporal unitary transformations. Using root-raised-cosine (RRC) shaped 16-QAM as an example, we show that high ($>30$ dB) signal-to-noise-and-distortion ratio (SINAD) is achievable if the bandwidth of the phase modulator is sufficient. We also reveal that the digital-to-analog converter quantization noise can cause an nonlinear increase in the overall system noise and distortion. Finally, by calculating the shot noise limited transmitter SINAD  of the unitary transformation scheme, we show that the phase modulation scheme can substantially outperform the lossy conventional approach at the same transmitter laser power.

\section{Distortions from finite stage spectro-temporal unitary transformations}


\begin{figure}[t]
   \centering
    \input{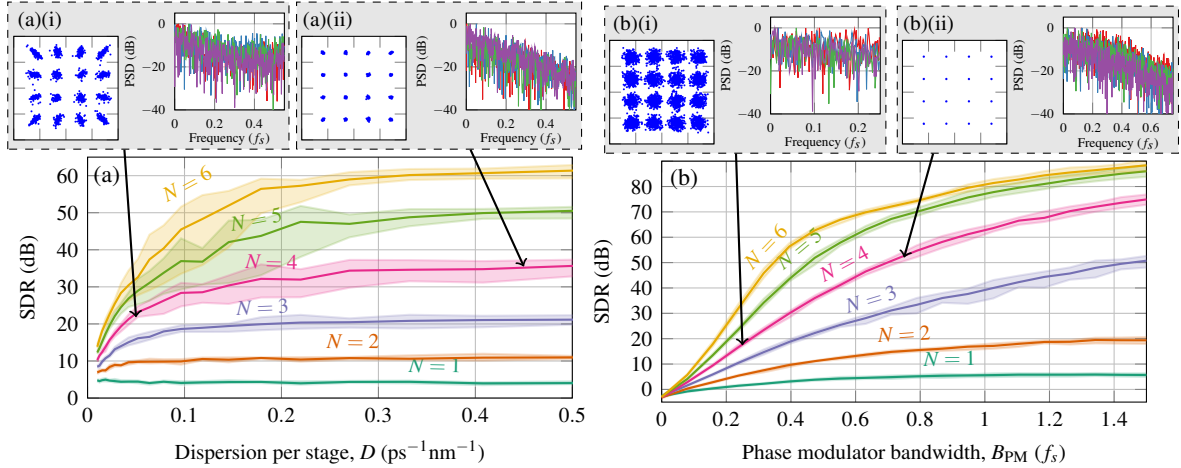}
       \vspace{-0.4cm}
    \caption{(a) SDR v. dispersion per stage ($D$) for a $B_{\textnormal{PM}}=$ 0.55~$f_s$, and varying number of stages, $N$. The dispersion is normalised to the symbol period squared ($T_s^2$), and so has units ps$^{-1}$nm$^{-1}$. (b) SDR v. phase modulator bandwidth ($B_{\textnormal{PM}}$), for $D=10~$ps$^{-1}$nm$^{-1}$, and varying number of stages, $N$. The phase modulator bandwidth is plotted in units of symbol rate, $f_s$. }
    \label{fig:single carrier}
\vspace{-0.6cm}

\end{figure}

To understand the performance of the spectro-temporal unitary transformation method scheme without loss of generality, we use an RRC ($\beta=0.1$) $16$-QAM signal to study the signal-to-distortion ratio (SDR). Here, the `distortions' are those introduced by the imperfect transformation matrix as limited by design constraints. We simulate 10 independent sets of 512 symbols, as 512 symbols is a suitable block length for real-time continuous operation when considering the computational complexity in solving the transformation matrix~\cite{saxena2023performance}. The required electrical waveforms for driving the phase modulations are calculated using a modified wavefront matching algorithm~\cite{sakamaki2007new}, in which the initial CW wave is propagated back and forth through the system model while minimising the phase difference at each stage. The phase update for the $n$-th phase modulator at iteration $k+1$ is given by \begin{equation}
    \phi_{n,k+1}(t) = \phi_{n,k}(t)-\mu \arg \Big[h(t)*F_n(t)B_n^*(t)e^{-i \phi_{n,k}(t)}\Big]
\end{equation} where $B_n(t)$ is the backwards propagating wave, $F_n(t)$ the forward propagating wave, $h(t)$ is the impulse repsonse of the phase modulator, and $\mu$ is the update step size. The algorithm is iterated until no improvements in SDR are seen. Although this not necessarily the computationally efficient algorithm~\cite{saxena2023performance}, it offers the most stable convergence to the target waveform~\cite{wyrowski1991upper} and so is useful for examining limits. The simulation is oversampled by a factor of eight (i.e. 8$f_s$) which ensures stable convergence. Fig.~\ref{fig:single carrier} shows the average converged symbol SDRs for the number of stages $N$ ($N=1,2,...,6$) while varying the dispersion per stage $D$ (Fig.~\ref{fig:single carrier}(a)) and the phase modulator bandwidth $B_{\textnormal{PM}}$ (Fig.~\ref{fig:single carrier}(b)). Shaded regions indicate the minimum and maximum SDRs of the 10 sequences. 

Fig.~\ref{fig:single carrier}(a) shows that, in general, the more number of stages available, the higher SDR can be achieved. At any given number of stages $N$, there is a maximum SDR even if $D$ can be increased indefinitely.  This is because the function of the dispersive element is to provide sufficient temporal mode mixing to approximate the required non-monomial matrix operation~\cite{borevich1981subgroups}, which is achieved at a finite amount of dispersion.  Overall, the distortion level can be very low (e.g. -35~dB) if a few (e.g. 4) stages with a small amount of dispersion per stage (e.g. 0.3~ps$^{-1}$nm$^{-1}$) can be implemented. For instance, at 200~GBd, 0.3~ps$^{-1}$nm$^{-1}$ is a dispersion of 7.5~ps/nm, which can easily be achieved on chip in standard silicon photonics~\cite{stern2023silicon}. In the extreme case of $N=1$, the SDR does not change with dispersion since in this case the dispersive efficiency of a single element is defined purely by the target waveform~\cite{wyrowski1991upper}. Our simulations showed that varying the block length, $\beta$, or QAM order does not substantially affect the converged SDR. This is expected since the dispersive efficiency of a single element is determined by the ratio $\langle| \Phi(t)|\rangle^2 / \langle| \Phi(t)|^2\rangle$ for target waveform $\Phi(t)$~\cite{wyrowski1991upper}, which does not vary substantially between these waveform types. The inset constellations highlight that when the amount of dispersion for a given $N$ is sub-optimal (Fig.~\ref{fig:single carrier}(a)(i)), the main source of noise is amplitude error, as is intuitive. On the contrary, when the SDR is limited by the number of stages (Fig.~\ref{fig:single carrier}(b)(i)), the noise appears equally distributed between phase and amplitude.

Fig.~\ref{fig:single carrier}(b) shows that increasing the phase modulator bandwidth allows for more accurate modulation, even for the case of $N=1$ which saturates at the dispersive limit for a single phase only element~\cite{wyrowski1991upper}. Excess phase modulator bandwidth therefore allows access to high (e.g. $>30$ dB) SDRs even with lower (e.g. $N=3$) modulator counts. The inset diagrams also plot some example spectrums of the converged phase modulations next to their corresponding constellations. One interesting observation is that the frequency content of the converged phase instructions in the two high SDR cases (Fig.~\ref{fig:single carrier}(a)(ii) and Fig.~\ref{fig:single carrier}(b)(ii)) shows a strong roll off, reaching around -20~dB at the maximum frequency. Therefore, increasing the bandwidth of the phase instructions may not actually significantly increase the 3~dB bandwidth requirements of the DAC/driver/modulator.


\section{Transmitter signal-to-noise-and-distortion ratio (SINAD)}

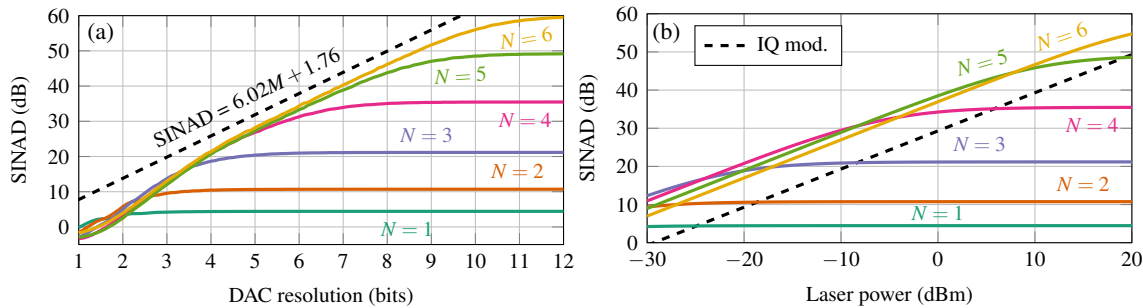
\begin{figure}[tb]
   \centering
 \pgfmathsetmacro{\figuresizex}{0.5}
\pgfmathsetmacro{\figuresizey}{0.29}

\begin{tikzpicture}[trim axis left,trim axis right]

\begin{axis} [ylabel= SINAD (dB), 
              xlabel={DAC resolution (bits)},
              xmin=1,xmax=12,
              ymax=60,
              ymin=-5,
              height=\figuresizey*\linewidth,
              grid=both,
              width=\figuresizex*\linewidth,
              cycle list/Dark2,
              ylabel near ticks,
              legend pos=north west,
              legend columns =3,
              clip mode=individual,
              log ticks with fixed point,
              legend style={font=\footnotesize},
              label style={font=\footnotesize},
              yticklabels={0,10,20,30,40,50,60,70},
              ytick={0,10,20,30,40,50,60,70},
              xticklabels={1,2,3,4,5,6,7,8,9,10,11,12},
              xtick={1,2,3,4,5,6,7,8,9,10,11,12},
              tick label style={font=\footnotesize},]

\addplot+ [mark=none, very thick] table [x=bits,y=1, col sep=comma] {data/dac_bits_v_SNR_seed_averaged.csv} node[pos=0.75, sloped,anchor=north]{{\footnotesize $N=1$}};

\addplot+ [mark=none, very thick] table [x=bits,y=2, col sep=comma] {data/dac_bits_v_SNR_seed_averaged.csv} node[pos=0.95, sloped,anchor=south]{{\footnotesize $N=2$}};

\addplot+ [mark=none, very thick] table [x=bits,y=3, col sep=comma] {data/dac_bits_v_SNR_seed_averaged.csv} node[pos=0.9, sloped,anchor=south]{{\footnotesize $N=3$}};

\addplot+ [mark=none, very thick] table [x=bits,y=4, col sep=comma] {data/dac_bits_v_SNR_seed_averaged.csv} node[pos=0.98, sloped,anchor=north]{{\footnotesize $N=4$}};

\addplot+ [mark=none, very thick] table [x=bits,y=5, col sep=comma] {data/dac_bits_v_SNR_seed_averaged.csv} node[pos=0.95, sloped,anchor=north]{{\footnotesize $N=5$}};

\addplot+ [mark=none, very thick] table [x=bits,y=6, col sep=comma] {data/dac_bits_v_SNR_seed_averaged.csv} node[pos=0.98, sloped,anchor=north]{{\footnotesize $N=6$}};

\addplot+[black, dashed,very thick, domain=0:10]{6.02*x + 1.76} node[pos=0.5, sloped,anchor=south]{{\footnotesize SINAD = $6.02M + 1.76$}};

\node at (axis cs:1.5, 55) {{\small (a)}};

\end{axis}

\end{tikzpicture}\hspace{1.1cm}\begin{tikzpicture}[trim axis left,trim axis right]

\begin{axis} [ylabel= SINAD (dB), 
              xlabel=Laser power  (dBm),
              xmin=-30,xmax=20,
              ymax=60,
              ymin=0,
              height=\figuresizey*\linewidth,
              grid=both,
              width=\figuresizex*\linewidth,
              cycle list/Dark2,
              ylabel near ticks,
              legend columns =3,
              clip mode=individual,
              log ticks with fixed point,
              legend style={font=\footnotesize, anchor=north,at={(0.25,0.95)}},
              label style={font=\footnotesize},
              yticklabels={0,10,20,30,40,50,60,70},
              ytick={0,10,20,30,40,50,60,70},
              tick label style={font=\footnotesize},]

\addplot+ [mark=none, very thick,black,dashed] table [x=laser_powers,y=IQ, col sep=comma] {data/laser_power_2dB_loss_0.3_Vp.csv} node[pos=0.32, sloped,anchor=north]{{}};
\addlegendentry{IQ mod.}

\addplot+ [mark=none, very thick] table [x=laser_powers,y=2, col sep=comma] {data/laser_power_2dB_loss_0.3_Vp.csv} node[pos=0.9, sloped,anchor=south]{{\footnotesize $N=2$}};

\pgfplotsset{cycle list shift=-2}

\addplot+ [mark=none, very thick] table [x=laser_powers,y=1, col sep=comma] {data/laser_power_2dB_loss_0.3_Vp.csv} node[pos=0.6, sloped,anchor=south,yshift=-2]{{\footnotesize $N=1$}};
\pgfplotsset{cycle list shift=-1}
\addplot+ [mark=none, very thick] table [x=laser_powers,y=3, col sep=comma] {data/laser_power_2dB_loss_0.3_Vp.csv} node[pos=0.7, sloped,anchor=south]{{\footnotesize $N=3$}};

\addplot+ [mark=none, very thick] table [x=laser_powers,y=4, col sep=comma] {data/laser_power_2dB_loss_0.3_Vp.csv} node[pos=0.93, sloped,anchor=north]{{\footnotesize $N=4$}};

\addplot+ [mark=none, very thick] table [x=laser_powers,y=5, col sep=comma] {data/laser_power_2dB_loss_0.3_Vp.csv} node[pos=0.75, sloped,anchor=south]{{\footnotesize $N=5$}};

\addplot+ [mark=none, very thick] table [x=laser_powers,y=6, col sep=comma] {data/laser_power_2dB_loss_0.3_Vp.csv} node[pos=0.88, sloped,anchor=south]{{\footnotesize $N=6$}};

\node at (axis cs:-28, 55) {{\small (b)}};

\end{axis}

\end{tikzpicture}
    \vspace{-0.4cm}

    \caption{(a) SINAD v. resolution of the DACs used to drive the phase modulators, for $B_{\textnormal{PM}}= 0.55~f_s$ and $D=0.3$~ps$^{-1}$nm$^{-1}$. (b) SINAD v laser power for 200 GBd, assuming a modulation loss per stage of 2~dB, $B_{\textnormal{PM}} =110$~GHz, $7.5~$~ps/nm dispersion. For the IQ mod. case, we assume a modulation depth of $V_{\textnormal{peak}}/V_\pi = 0.3$.}
    \label{fig2}
   \vspace{-0.8cm}

\end{figure}

Due to the nonlinear relationship between the successive phase modulation stages, the impact of the quantization noise from the digital-to-analog converters (DACs) is not as straightforward as MZM-based modulations. Therefore, we quantize the previously calculated phase profiles to add quantization noise and examine the impact of DAC resolution. We use an example of $B_{\textnormal{PM}} =0.55~f_s$, $D = 0.3~$ps$^{-1}$nm$^{-1}$ and plot the overall SINAD in Fig.~\ref{fig2}(a). At high resolution, the SINAD simply saturates at the transform limited SDRs presented in Fig.~\ref{fig:single carrier}. However, at low resolution, the SINAD is substantially lower than the theoretical limit ($6.02M + 1.76$, for $M$ bits), which is a result of the nonlinear relationship between the successive phase modulation stages. Nevertheless, Fig.~\ref{fig2}(a), shows that a practical DAC resolution of 5 or 6 bits would offer a sufficiently low noise/distortion floor (e.g. $<-$30 dB). 

Although the spectro-temporal unitary transformation is theoretically lossless, the phase modulators and dispersive elements will in practice have insertion loss. Therefore, in Fig.~\ref{fig2}(b) we calculate the transmitter SINAD vs. on-chip laser power at the modulator input for a given number of stages, $N$, considering shot noise (i.e. optical signal to noise ratio) and the previously calculated transformation distortions. We use a practical example of 200~GBd symbol rate (i.e. 200~GHz transmitter optical bandwidth), with $B_{\textnormal{PM}} =110$~GHz, $7.5~$~ps/nm dispersion and assume each stage has a realistic 2-dB insertion loss~\cite{stern2023silicon}. We also calculate modulation loss for the conventional IQ case from the actual Nyquist-shaped 16-QAM electrical waveform driving the sinusoidal MZM transfer function with a modulation depth of $V_{\textnormal{peak}}/V_\pi = 0.3$. The modulation depth is chosen for linear operation~\cite{chen201916384}, resulting in an average modulation loss of 21~dB. At all laser powers, Fig.~\ref{fig2}(b) shows that the the low loss nature of the transformation allows it to achieve higher SINADs than the conventional IQ modulation case, provided enough stages are implemented. Since fewer stages results in lower modulation loss, the performance gain is most dramatic at low laser powers, highlighting the potential for low power or high bandwidth ($>500$~GHz) transmitters.

\section{Conclusion}

We have analysed the performance limits of spectro-temporal unitary transformations for coherent modulation and shown that they can achieve high signal-to-noise-and-distortion ratios with a low, practical number of phase modulators and a dispersion that is achievable on chip. We also highlight that they can substantially outperform conventional IQ modulation in terms of achievable SINAD for a given laser power due to their low loss nature.
   \vspace{-0.2cm}

\bibliographystyle{opticajnl}
\bibliography{sample}

\begin{thebibliography}{10}
\newcommand{\enquote}[1]{``#1''}

\bibitem{mazur2019optical}
M.~Mazur \emph{et~al.}, \enquote{Optical arbitrary waveform generator based on time-domain multiplane light conversion,} in \emph{OFC,}  (2019), p. M1B.3.

\bibitem{thiel2017programmable}
V.~Thiel \emph{et~al.}, \enquote{Programmable unitary transformation of spectro-temporal modes,} in \emph{Laser Science,}  (2017), pp. JW4A--5.

\bibitem{morizur2010programmable}
J.-F. Morizur \emph{et~al.}, \enquote{Programmable unitary spatial mode manipulation,} {\protect\JournalTitle{JOSA A}} \textbf{27}, 2524--2531 (2010).

\bibitem{saxena2023performance}
B.~Saxena \emph{et~al.}, \enquote{Performance comparison between all-pass and {IQ} optical modulation,} {\protect\JournalTitle{Journal of Lightwave Technology}} \textbf{42}, 201--207 (2023).

\bibitem{mazur2019multi}
M.~Mazur \emph{et~al.}, \enquote{Multi-wavelength arbitrary waveform generation through...} {\protect\JournalTitle{arXiv preprint arXiv:1907.02595}}  (2019).

\bibitem{ashby2020temporal}
J.~Ashby \emph{et~al.}, \enquote{Temporal mode transformations by sequential time and frequency phase...} {\protect\JournalTitle{Optics Express}} \textbf{28}, 38376--38389 (2020).

\bibitem{sakamaki2007new}
Y.~Sakamaki \emph{et~al.}, \enquote{New optical waveguide design based on wavefront matching method,} {\protect\JournalTitle{JLT}} \textbf{25}, 3511--3518 (2007).

\bibitem{wyrowski1991upper}
F.~Wyrowski, \enquote{Upper bound of the diffraction efficiency of diffractive phase elements,} {\protect\JournalTitle{Optics letters}} \textbf{16}, 1915--1917 (1991).

\bibitem{borevich1981subgroups}
Z.~Borevich \emph{et~al.}, \enquote{Subgroups of the unitary group that contain the group of diagonal matrices,} {\protect\JournalTitle{Journal of Soviet Mathematics}} \textbf{17}, 1951--1959 (1981).

\bibitem{stern2023silicon}
B.~Stern \emph{et~al.}, \enquote{Silicon photonic direct-detection phase retrieval receiver,} in \emph{ECOC 2023,}  (2023), p. M.A.2.2.

\bibitem{chen201916384}
X.~Chen \emph{et~al.}, \enquote{{16384-QAM} transmission at 10 {GBd} over 25-km {SSMF} using polarization-...} in \emph{ECOC 2019,}  (2019), p. PD3.3.

\end{thebibliography}



\end{document}